\documentstyle{article}\begin{document}
\title{Energy and entropy of relativistic diffusing particles }
\author{ Z. Haba\\
Institute of Theoretical Physics, University of Wroclaw,\\ 50-204
Wroclaw, Plac Maxa Borna 9,
Poland\\PACS:05.10.Gg,05.20.Dd,98.80.Jk}
\date{\today}\maketitle
\begin{abstract}
We discuss energy-momentum tensor and the second law of
thermodynamics for a system of relativistic diffusing particles. We
calculate the energy and entropy flow in this system. We obtain an
exact time-dependence of energy, entropy and free energy of a beam
of photons in a reservoir of a fixed temperature.
\end{abstract}
\section{Introduction}
 A probabilistic description  of a particle in an
 environment of some other particles is well-known in the
 non-relativistic classical and quantum mechanics \cite{pitaj}. The criteria
 for a diffusive approximation of such systems have been discussed
  for a long time. The diffusive approximation determines a single
  particle probability distribution. Then, we may consider a
  stream of particles with the same probability distribution and apply methods of statistical mechanics
  for a thermodynamic or hydrodynamic description of such a system.
  This is the well-known model of non-equilibrium thermodynamics \cite{mazur}\cite{glans}.

  A relativistic diffusion approximation is less developed (for a review see \cite{dunkel}\cite{debrev}).
  There are prospects for applications mainly in heavy ion collisions, plasma physics
  (ordinary electron-ion plasma and quark-gluon plasma) and in astrophysics \cite{rybicki}.
   There
  are  obstacles in such an approach
   because  relativistic dynamics of
  multiparticle systems  encounters some conceptual problems.
    Quantum field theory gives a relativistic description of
  scattering processes. However, it  encounters difficulties
  with a notion of interacting particle systems at a finite time.
  Nevertheless, if we interpret the Wigner function as the probability distribution then
    relativistic transport equations can be derived
  from quantum field theory \cite{zachar}\cite{groot}. A diffusion approximation to such
  equations has been applied in high energy physics
  \cite{hwa}\cite{svet}\cite{rafelski}\cite{ion}.
  In astrophysics a diffusion approximation to a description of light moving
  through a space  filled with an electron gas has found  wide
  spread applications \cite{rybicki}.

  Putting aside the problem of a derivation of a diffusion equation from multiparticle dynamics we can
  ask the question about its mathematical form if it is to satisfy the postulates of the relativistic mechanics.
  The mathematical problem of a relativistic diffusion has been
  posed by Schay \cite{schay} and Dudley \cite{dudley}. It comes
  out that if the particle mass is defined as the square of the four-momentum and the diffusion
  is to be a Markov process then the answer is unique. The
  generator of the diffusion must be defined on the mass-shell as the second order
  relativistic invariant differential operator.
  We usually are interested in systems which are close to
  equilibrium (otherwise a mathematical description would not be
  feasible). The notion of an equilibrium is  relativistically
  covariant but not invariant. If we assume the usual postulates
  of  equilibrium statistical mechanics  then the
  reversible relativistic Markov process is uniquely determined by
  the detailed balance principle. We have discussed relativistic
  diffusions with an equilibrium distribution in \cite{habapre}\cite{habajmp}.

 Many particle systems can also be described approximately by relativistic hydrodynamics.
 The hydrodynamics can in principle be derived from mechanics or through an intermediate approximation
 step from diffusive dynamics. The
  form of the energy-momentum tensor (as expressed by fluid velocity)
    constitutes the  basic assumption of the relativistic
    hydrodynamics \cite{landau}\cite{israel}\cite{rom}.
We discuss the   energy-momentum tensor in diffusion theory
(sec.3) but do not attempt  to derive its hydrodynamic form as
expressed by fluid velocity. The energy-momentum tensor is defined
by the
  particle's probability distribution on the phase space.
  It  is not
  conserved because the energy of a stream of particles is
  dissipated in a medium.   The diffusion
  equations determine the time evolution of the energy-momentum
  tensor. The definition of the probability distribution and an
  equilibrium  distribution allows to define the relative entropy
  (sec.4)(the relative entropy for a class of relativistic diffusions is discussed also in
  \cite{debentropy}\cite{deb2}) which in the theory of non-relativistic diffusions has been
   related to free energy \cite{haken} and subsequently to
    the entropy
  and internal energy.  The relation of the relative entropy to the thermodynamics of
  non-relativistic diffusing systems has  been discussed in \cite{haken}
  \cite{nicolis}\cite{bag}. In this letter we show such a relation for
  relativistic diffusions. We obtain equations for a time
  evolution of the energy and entropy. The case of a diffusion
  of massless particles is exactly soluble. We have shown \cite{habampa} that
  the diffusion of massless particles can be considered as a
  linear
  approximation to the non-linear Kompaneets diffusion \cite{komp}.
 In this letter (sec.5) we obtain exact  results for a time evolution of  energy and
entropy  of radiation within a linear relativistic diffusion
theory.
\section{The relativistic diffusion
} Following \cite{schay}\cite{dudley}\cite{habapre} we begin with
a proper time evolution of a function $\phi$ on the phase space

\begin{equation}
\partial_{\tau}\phi={\cal G}\phi=(p^{\mu}\partial_{\mu}^{x}+{\cal
A})\phi,
\end{equation}
where ${\cal A}$ is the second order differential operator defined
on the mass-shell \begin{displaymath}
p_{\mu}p^{\mu}=p_{0}^{2}-{\bf p}^{2}=m^{2}c^{2}.
\end{displaymath}Differentiation over space-time coordinates
has an index $x$ whereas differentiation without an index concerns
momenta, space-time indices are denoted by Greek letters whereas
spatial indices by Latin letters. The probability density $\Phi$
evolves according to an adjoint equation
\begin{equation}
\partial_{\tau}\Phi_{\tau}={\cal
G}^{*}\Phi_{\tau}=(-p^{\mu}\partial_{\mu}^{x}+{\cal
A}^{*})\Phi_{\tau}
\end{equation} resulting from
\begin{equation}
\int dxd{\bf p}\phi_{\tau}(x,{\bf p})\Phi(x,{\bf p})= \int dxd{\bf
p} \phi(x,{\bf p})\Phi_{\tau}(x,{\bf p}). \end{equation} The
probability density $\Phi$ of a particle in a laboratory frame is
independent of $\tau$ and (from eq.(2)) is the solution of the
equation
\begin{equation}
{\cal G}^{*}\Phi=0.
\end{equation}
 If
we choose the spatial momenta ${\bf p}$ as coordinates on the
mass-shell then the $O(3,1)$ invariant diffusion generator reads

\begin{equation}\begin{array}{l}
2\gamma^{-2}{\cal A}_{0}=\triangle_{H}=
(\delta_{jk}+m^{-2}c^{-2}p_{j}p_{k})\frac{\partial}{\partial
p_{j}}\frac{\partial}{\partial
p_{k}}+3m^{-2}c^{-2}p_{k}\frac{\partial}{\partial p_{k}}.
\end{array}\end{equation}
We shall also use $\kappa^{2}=m^{-2}c^{-2}\gamma^{2}$ as the
diffusion constant. A diffusion generated by ${\cal A}_{0}$ does not
have a finite equilibrium measure. In order to achieve an
equilibrium we add a friction term $R=R_{j}\partial^{j}$  to ${\cal
A}_{0}$. Now,
\begin{equation}
{\cal A}={\cal A}_{0}+R_{j}\partial^{j}\equiv
A_{jk}\partial^{j}\partial^{k}+B_{j}\partial^{j},
\end{equation}
where the drift $B$ is
\begin{equation}B_{k}\equiv R_{k}+\frac{3\kappa^{2}}{2}p_{k}.
\end{equation}
In an electromagnetic field we still add the force
\begin{equation}
G=\frac{e}{mc}F_{j\nu}p^{\nu}\partial^{j},
\end{equation}(where $e$ is an electric charge and $F_{\mu\nu}$ is the electromagnetic field strength tensor).

We assume that there exists a time independent equilibrium
solution $\Phi_{E}$ of eq.(4). If the Markov process is reversible
\cite{kol} then the probability distribution $\Phi$ and the
invariant measure $\Phi_{E}$ satisfy the detailed balance
condition \cite{haken} which determines the drift $R$ as a
function of the invariant measure. Applying basic assumptions of
the classical relativistic statistical mechanics we can conclude
that the equilibrium distribution of free relativistic particles
is (the J\"uttner distribution \cite{juttner})
\begin{equation}
\Phi_{E}=p_{0}^{-1}\exp(-\beta cp_{0}).
\end{equation}
where $\beta^{-1}=k_{B}T$, $k_{B}$ is the Boltzmann constant and
$T$ is the temperature.
 $\Phi_{E}$ determines the drift
\begin{equation} R_{j}=-\frac{\kappa^{2}}{2}p_{j}\beta
cp_{0}.
\end{equation}
We can also find an equilibrium measure of a particle in an
electric field. Under the assumption that only the spatial
components $F_{0j}=-c\partial^{x}_{j}V$ of $F_{\mu\nu}$ are
different from zero we have\begin{equation}
\Phi_{E}^{V}=p_{0}^{-1}\exp(-\beta (cp_{0}+eV)).
\end{equation}
Then, the drift is
\begin{equation} R_{j}^{V}=-\frac{\kappa^{2}}{2}p_{j}\beta
cp_{0} +\frac{e}{mc}p_{0}\partial_{j}^{x}V
\end{equation}
(it could be obtained by an addition of the force (8) to the
friction (10)).

The limit $m\rightarrow 0$ is particularly simple. It can be
obtained as the limit $m\rightarrow 0$ of the time evolution
$\exp(\tau m^{2}{\cal G})$ \cite{habampa}. Then, eq.(4) reads
\begin{equation}\begin{array}{l}
\vert {\bf
p}\vert\partial_{t}\Phi=\frac{c\kappa^{2}}{2}\partial^{j}\partial^{k}p_{j}p_{k}\Phi
-\frac{3c\kappa^{2}}{2}\partial^{j}p_{j}\Phi+\frac{\kappa^{2}}{2}\beta
c^{2}\partial^{j}p_{j}\vert{\bf p}\vert\Phi+c{\bf p}\nabla_{x}\Phi
\end{array}\end{equation}We write
\begin{equation} \Phi=\Phi_{E}\Psi
\end{equation}
 and $r=\vert {\bf p}\vert$. Then, the equation for $\Psi(\vert{\bf p}\vert,{\bf n},{\bf x})$
 is related to the Bessel diffusion \cite{ikeda}

\begin{equation}
\partial_{t}\Psi=\frac{c\kappa^{2}}{2} r\partial_{r}^{2}\Psi+(\frac{3c\kappa^{2}}{2}
-\frac{\beta c^{2}\kappa^{2}}{2} r)\partial_{r}\Psi -c{\bf
n}\nabla_{{\bf x}}\Psi,
\end{equation}
where ${\bf n}={\bf p}\vert{\bf p}\vert^{-1}$ is a fixed vector.
 \section{Energy-momentum tensor}

We define a configuration space density
\begin{equation}
\rho(x)=\int d{\bf p}\Phi(x,{\bf p}).
\end{equation}
The particle  density current is defined by
\begin{equation}
N^{\mu}=\int d{\bf p}p^{\mu}\Phi(x,{\bf p}) .\end{equation} From
the transport equation (4) we derive the current conservation
\begin{equation}
\partial_{\mu}N^{\mu}=0.
\end{equation}
As a consequence
\begin{equation}Z= \int d{\bf x} N^{0}=\int d{\bf x}d{\bf
p}p_{0}\Phi=const.
\end{equation}
We  we can apply eq.(19) to normalize the probability distribution.

We define the  energy-momentum tensor \cite{groot}
  \begin{equation}
 T_{\mu\nu}(x)=\int d{\bf p}p_{\mu}p_{\nu}\Phi({\bf x},{\bf p}).
 \end{equation}
 The energy-momentum tensor gives a covariant description of
 a continuous distribution of the energy and momentum.
 $T_{0\mu}$ has the meaning of the four-momentum density current.
 Then
 \begin{displaymath}
 {\cal P}_{\mu}=\int d{\bf x}T_{0\mu},
 \end{displaymath}
 determines the four-momentum of a stream of particles. We can write
\begin{equation}
T_{0\mu}=\int d{\bf p}\Omega({\bf p},{\bf x})p_{\mu}
\end{equation} where the physical meaning of $T_{0\mu}$ and ${\cal P}_{\mu}$
imply that $ \Omega({\bf p},{\bf x})=\Phi({\bf p},{\bf x}) p_{0} $
is the phase space probability distribution (after a
normalization).

We obtain simple equations for the divergence of the kinetic
energy-momentum in an electromagnetic field at $\beta=0$ (no
friction). Then,
\begin{equation}
\partial^{\mu}T_{\mu\nu}=
 \frac{3\kappa^{2}}{2}N_{\nu}+
\frac{e}{mc}F_{\nu\sigma}N^{\sigma}.
\end{equation}
If friction is present then we consider a diffusion in the electric
field $E_{k}=-\partial_{k}V$ ( in such a case eq.(4) has an
equilibrium solution (11)). Then, from eq.(4)
\begin{equation}
\partial^{\mu}T_{\mu 0}=\frac{3\kappa^{2}}{2}N_{0}-\frac{\kappa^{2}}{2}
\beta c T_{00}+\frac{\kappa^{2}}{2}\beta
m^{2}c^{3}\rho+\frac{e}{m}\partial_{k}V N_{k}
\end{equation}and
\begin{equation}
\partial^{\mu}T_{\mu k}=
 \frac{3\kappa^{2}}{2}N_{k}-\frac{\kappa^{2}}{2}
\beta cT_{0k}+ \frac{e}{mc}\partial_{k}V N^{0}.
\end{equation}
Eqs.(23)-(24) say that the flow of energy and momentum is determined
by the currents.

The kinetic energy can be defined as
\begin{equation}
{\cal W}=c{\cal P}_{0}=c\int d{\bf x} T_{00}=c\int d{\bf x} d{\bf
p}\Phi p_{0}^{2}.
\end{equation}Then, from eq.(23)\begin{equation}
\partial_{0}{\cal W}=\frac{3\kappa^{2}c}{2}Z^{V}-\frac{\kappa^{2}}{2}
\beta c {\cal W}+\frac{\kappa^{2}}{2}\beta m^{2}c^{4}\int d{\bf
x}\rho+\frac{ec}{m}\int d{\bf x}\nabla V{\bf N}
\end{equation} where the normalization constant (19) in the case
of a potential $V$ is denoted $Z^{V}$.

It follows from eq.(26) that if there is no friction ($\beta=0$)
and no potential ( $V=0$) then the energy grows linearly in time.
If we include the external potential $V$ into the energy
\begin{equation}{\cal W}_{V}= \int d{\bf
x}(cT_{00}+ep_{0}V\Phi)\equiv \int d{\bf p}\Omega({\bf
p})(cp_{0}+eV)\equiv {\cal W}+e\langle V\rangle.\end{equation}
then eq.(26) reads
\begin{equation}
\partial_{0}W_{V}=\frac{3}{2}\kappa^{2}cZ^{V}-\frac{\kappa^{2}}{2}
\beta c {\cal W}+\frac{\kappa^{2}}{2}\beta m^{2}c^{4}\int d{\bf
x}\rho.\end{equation} We show that the energy  ${\cal W}_{V}$ is
bounded if $V$ is bounded and $\beta>0$. Let
\begin{equation}
\nu=\beta\kappa^{2}c^{2}.
\end{equation}
We  write eq.(28) in an integral form
\begin{equation}\begin{array}{l}
{\cal W}_{V}(t)=\exp(-\frac{\nu}{2} t){\cal W}_{V}(t=0)\cr +
\frac{\kappa^{2}c^{2}}{2}\exp(-\frac{\nu }{2}t) \int_{0}^{t}
\exp(\frac{\nu}{2} s)(3Z^{V}+\beta c^{3}m^{2}\int d{\bf
x}\rho+e\beta \langle V\rangle) ds.\end{array}\end{equation} Then,
from eq.(28) and the inequality
\begin{displaymath}\int d{\bf x}\rho\leq \int d{\bf p}d{\bf
x}(mc)^{-1}p_{0}\Phi=(mc)^{-1}Z^{V}\end{displaymath} we have
\begin{equation}
 {\cal W}_{V}\leq \exp(-\frac{\nu}{2} t){\cal
W}_{V}(t=0)+\beta^{-1}\Big(1- \exp(-\frac{\nu}{2}
t)\Big)Z^{V}(3+\beta mc^{2}+e\beta v ),
\end{equation} where $v=sup\vert V\vert$.
\section{Relative
entropy, entropy and free energy} The relative entropy (also
called Kullback-Leibler entropy) determines a distance between two
probability measures  . Define the relative entropy of two
unnormalized probability distributions $\Phi$ and $\Phi_{E}$ ($Z$
and $Z_{E}$ are the normalization constants) as
\begin{equation}
S_{K}(\Phi;\Phi_{E})=Z^{-1}\int d{\bf x}d{\bf p}p_{0}\Phi\ln\Big(
Z^{-1}\Phi(\Phi_{E})^{-1}Z_{E}\Big),\end{equation} where
\begin{equation}
Z_{E}=\int d{\bf x}d{\bf p}\exp(-\beta cp_{0})
\end{equation}
(the system must be in a finite volume if $Z_{E}$ is to be
finite). It is known that \cite{risken}
\begin{equation} S_{K}(\Phi;\Phi_{E})\geq 0.
\end{equation}
An easy calculation using the transport equation (4) gives
\begin{equation}\begin{array}{l}
\partial_{0}S_{K}(\Phi;\Phi_{E})
=-Z^{-1}\int d{\bf p}d{\bf x}\Phi A_{jk}\partial_{j}\ln
Q\partial_{k}\ln Q\leq 0,
\end{array}\end{equation} where
\begin{equation}
Q=\Phi(\Phi_{E})^{-1}
\end{equation}
and $A_{jk}$ is defined in eq.(6). Eq.(35) holds true for any two
probability distributions solving the same Fokker-Planck equation
\cite{risken}.
 It follows from eqs.(34)-(35) that
$S_{K}(\Phi;\Phi_{E})$ is a non-negative function monotonically
decreasing to zero  at the equilibrium $\Phi_{E}$ (however, the
gradients in eq.(35) do not depend on the potential $V$). For
non-relativistic diffusions the properties of the relative entropy
are known for a long time
\cite{haken}\cite{risken}\cite{leb}\cite{lebo} (the decrease of
the relative entropy for a class of relativistic diffusions has
been shown in \cite{debentropy}\cite{deb2}).

The relative entropy in an electric field is
\begin{equation}\begin{array}{l}
S_{K}(\Phi;\Phi_{E}^{V})=Z^{-1}\int d{\bf x}d{\bf
p}p_{0}\Phi\ln\Big(Z^{-1}\Phi Z_{E}^{V}\exp(\beta(
cp_{0}+eV))\Big).
\end{array}\end{equation}
Here
\begin{equation}
Z_{E}^{V}=\int d{\bf x}d{\bf p}\exp(-\beta (cp_{0}+eV)).
\end{equation}
If we have a single particle distribution $\Phi$ then we can
define the entropy current
\begin{equation}
S^{\mu}(\Phi)=-k_{B}\int \frac{d{\bf
p}}{p_{0}}p^{\mu}(p_{0}\Phi)\ln\Big( p_{0}\Phi\Big)
\end{equation}
and  the Boltzmann entropy
\begin{equation} S(\Phi)=\int d{\bf x} S^{0}(\Phi).\end{equation}
From the definitions (25),(32) and (40) we obtain the relation
\begin{equation} ZS_{K}(\Phi;\Phi_{E})=-k_{B}^{-1}S+\beta {\cal
W}+Z\ln
 (Z_{E}Z^{-1})
 \end{equation}
We define the free energy
\begin{equation} {\cal
F}=\beta^{-1}ZS_{K}(\Phi;\Phi_{E})-Z\beta^{-1}\ln (ZZ_{E}^{-1})
\end{equation}and in a potential $V$
\begin{equation} {\cal
F}_{V}=\beta^{-1}Z^{V}S_{K}(\Phi;\Phi
_{E}^{V})-(Z^{V})^{-1}\beta^{-1}\ln (Z^{V}(Z_{E}^{V})^{-1})
\end{equation}
Then, from eqs.(25),(40) and (42) we obtain the basic
thermodynamic relation
\begin{equation}TS={\cal W}-{\cal F}.
\end{equation}

 $Z$ has the
meaning of the mean number of modes. We can define the energy per
mode
\begin{equation} w=Z^{-1}{\cal W}
\end{equation}
$ f=Z^{-1}{\cal F}$  and $s= Z^{-1}{\cal S}$. Then, the
thermodynamic equality (44) reads \begin{equation} Ts=w-f
\end{equation}
 We
can calculate the time derivative of the entropy
\begin{equation}\begin{array}{l}
T\partial_{0}S=\partial_{0}{\cal W}-\partial_{0}{\cal F}\cr
=\frac{3c\kappa^{2}}{2}Z-\frac{\kappa^{2}}{2} \beta c {\cal
W}+\frac{\kappa^{2}}{2}\beta m^{2}c^{4}\int d{\bf x}\rho+\int
d{\bf p}d{\bf x}\Phi A_{jk}\partial_{j}\ln Q\partial_{k}\ln
Q\end{array}\end{equation} ($Q=\Phi\Phi_{E}^{-1}$).The negative
contribution to the change of entropy comes from the energy loss
into the surroundings caused by the friction in the medium (see
the discussion of the entropy balance in
\cite{mazur}\cite{glans}).
 The question whether the entropy is increasing or not depends on the
relative strength of the terms on the rhs of eq.(47). The last
term on the rhs of eq.(47) tends to zero when the equilibrium is
approached. The sum of first three terms also tends to zero. In
order to prove this we take the equilibrium limit on the rhs of
eq.(47) (here $z=\int d{\bf p}\exp(-\beta cp_{0})$)
\begin{equation}
\begin{array}{l}
lim_{t\rightarrow\infty}\kappa^{-2}\partial_{0}S \cr
=\frac{3c}{2}z-\frac{1}{2} \beta c \int d{\bf p}p_{0}\exp(-\beta
cp_{0})+\frac{1}{2}\beta m^{2}c^{4}\int d{\bf p}
p_{0}^{-1}\exp(-\beta cp_{0})=0\end{array}\end{equation} The
result (48) comes from direct calculations  of the integrals (with
a use of some identities for modified Bessel functions $K_{\nu}$).
We can see that at large time there is a subtle cancellation of
the first three terms with the last one on the rhs of eq.(48).
Such an entropy balance is characteristic for a system exchanging
energy (and entropy) with the surroundings of the system until it
achieves an equilibrium \cite{mazur}\cite{glans}. The problem
whether the entropy ${\cal S}$ grows or not may depend on the
initial conditions, on the temperature and on time. If we choose
the initial condition $\Phi$ so that the initial ${\cal W}$ is
small and far from its equilibrium value then the rhs of eq.(47)
will be positive for a small time and will continue to be positive
until it approaches zero when $t\rightarrow\infty$ and
$\Phi\rightarrow\Phi_{E}$. In the opposite case of large initial
${\cal W}$ the entropy begins to decrease at $t=0$ and it may
continue doing so as time goes on. In the Appendix we consider an
example of a beam of particles prepared at $t=0$ in the
equilibrium state of temperature $T^{\prime}$. We calculate the
rhs of eq.(47) at $t=0$. If $T^{\prime}>T$ the energy and entropy
start to decrease (although $-{\cal F}$ grows) because of the
dissipation of the energy in the reservoir. In the opposite case
$T^{\prime}<T$ the beam acquires the energy and the entropy from
the reservoir (hence, $\partial_{0}{\cal S}>0$).   The beam can
gain the entropy till it achieves the equilibrium at the
temperature $T$. In the next section we prove that this is exactly
so in a soluble model of diffusing photons.

 We could  calculate the time derivative of the
entropy  directly from the definition (40)
\begin{equation}
\partial_{0}S(\Phi)=-k_{B}\int\Big(({\bf p}\nabla+ {\cal A }^{*})\Phi\Big)\ln(p_{0}\Phi)-k_{B}\int ({\bf
p}\nabla+ {\cal A }^{*})\Phi.
\end{equation}
Eq.(49) shows (as  ${\cal A}^{*}\Phi_{E}=0$) that
\begin{equation}
\partial_{0}S(\Phi_{E})=0
\end{equation}
(no entropy production in an equilibrium). The conclusion (50) is
in agreement with (48) because $\Phi\rightarrow \Phi_{E}$ as
$t\rightarrow \infty$.

\section{Thermodynamics
 of diffusing photons} We have shown in
\cite{habampa} \cite{habajpa} that a diffusion of photons can be
described by the limit $m\rightarrow 0$ of the diffusion equation.
This equation can be solved explicitly. We write
$\Phi=\Phi_{E}\Psi$. We choose the initial condition $\Psi$ for
eq.(15) in the form of the Laplace transform (${\bf n}={\bf
p}\vert{\bf p}\vert^{-1}$)
\begin{equation}\Psi({\bf x},{\bf p})= \Psi({\bf x},\vert{\bf p}\vert,{\bf n})
=\int_{0}^{\infty}d\sigma \tilde{\Psi}({\bf x},\sigma,{\bf n})
\exp(-\sigma c\vert{\bf p}\vert).\end{equation} First, we choose
$\tilde{\Psi}=\delta(\sigma-\lambda)$. Let
\begin{equation} A(\lambda,t)=\Big(1
+\lambda\beta^{-1}(1-\exp(-\frac{\nu t}{2}))\Big),
\end{equation}then the solution of
eq.(15) with the initial condition $\exp(-c\lambda \vert {\bf
p}\vert)$ is
\cite{ikeda}\cite{habajmp}\begin{equation}\begin{array}{l}\Psi_{t}^{\lambda}({\bf
p})=A(\lambda,t)^{-3}
\exp\Big(-A(\lambda,t)^{-1}c\lambda\exp(-\frac{\nu}{2} t)\vert{\bf
p}\vert\Big).\end{array}
\end{equation}In general,
\begin{equation}
\Psi_{t}({\bf x},\vert{\bf p}\vert,{\bf n})
=\int_{0}^{\infty}d\sigma \tilde{\Psi}({\bf x}-{\bf
n}ct,\sigma,{\bf n}) \Psi_{t}^{\sigma}({\bf p}).\end{equation}
From eqs.(52)-(54) we obtain the limit
\begin{equation}
\begin{array}{l}\lim_{t\rightarrow \infty}\int d{\bf x}d{\bf p}\Phi_{t}({\bf
x},{\bf p})M({\bf p})=\int d{\bf p}\Phi_{E}M({\bf p})\cr\Big(\int
d{\bf x}d{\bf p} \Psi({\bf x},{\bf p})\exp(-c\beta \vert{\bf
p}\vert)\Big)\Big( \int d{\bf p}\exp(-c\beta \vert{\bf
p}\vert)\Big)^{-1}.\end{array}\end{equation} Let
\begin{equation}
M({\bf p})=p_{0}H({\bf p}).
\end{equation}
Then, eq.(55) can be expressed in a form which looks like a
standard limit of mean values \begin{equation}
\begin{array}{l}\lim_{t\rightarrow \infty}\int d{\bf x}d{\bf p}\Phi_{t}({\bf
x},{\bf p})p_{0}H({\bf p})\Big(\int d{\bf x}d{\bf
p}\Phi_{t}p_{0}\Big)^{-1}\cr=\int d{\bf p}\Phi_{E}p_{0}H({\bf
p})\Big(\int d{\bf
p}p_{0}\Phi_{E}\Big)^{-1}.\end{array}\end{equation}
 We can  study the time
dependence of thermodynamic functions explicitly using the explicit
solution of the diffusion equation. When $m=0$ and $V=0$ then  we
obtain a closed equation (28) for ${\cal W}$ with the solution
\begin{equation}\begin{array}{l}
{\cal W}(t)=\exp(-\frac{\nu}{2} t){\cal W}(t=0)+
3Z\beta^{-1}(1-\exp(-\frac{\nu}{2}  t)).
\end{array}\end{equation}
We study in more detail the case
\begin{equation}
\Phi^{\lambda}=\Phi_{E}^{\prime}=p_{0}^{-1}\exp(-\beta^{\prime}cp_{0})=\Phi_{E}
\exp(-c\lambda\vert{\bf p}\vert),
\end{equation}
(in a finite volume which we set  as $1$) where
\begin{equation}
\lambda=\beta^{\prime}-\beta.
\end{equation}
The state $\Phi^{\lambda}$ describes a beam of photons coming from
a source of a fixed temperature $T^{\prime}$. We are interested in
a time evolution of this beam when passing through a medium of a
fixed temperature $T$. We assume that the interaction of  photons
with the medium can be described approximately as the diffusion.
Such an approximation has been established by Kompaneets in a
model of an interaction of photons with an electron gas
\cite{komp}. In eq.(58) for the model (59) we have
\begin{equation}
{\cal W}(t=0)=c\int d{\bf p}p_{0}^{2}\Phi_{E}^{\prime}=24\pi
c(c\beta+c\lambda)^{-4}=24\pi c^{-3}\beta^{\prime -4}
\end{equation}
(the Stefan-Boltzmann law) and
\begin{equation}
Z=\int d{\bf p}p_{0}\Phi_{E}^{\prime}=8\pi
(c\beta+c\lambda)^{-3}=8\pi c^{-3}\beta^{\prime -3}.
\end{equation}
Hence
\begin{equation}
{\cal W}(t)=24\pi c^{-3}\beta^{-1}\beta^{\prime-3}-24\pi \lambda
\exp(-\frac{\nu}{2}t)c^{-3}\beta^{-1}\beta^{\prime-4}
\end{equation}
and
\begin{equation}\begin{array}{l} \beta^{\prime
4}(24\pi)^{-1}Zc^{3}S_{K}(\Phi_{E}^{\prime},\Phi_{E})=-\beta^{\prime}\ln\Big(1+\lambda
\beta^{-1}(1-\exp(-\frac{\nu}{2}t))\Big) -\lambda
\exp(-\frac{\nu}{2}t).
\end{array}
\end{equation}
 Hence,
 \begin{equation}
\partial_{t}{\cal F}=-12\pi c^{-3}\lambda^{2}\nu\beta^{-1}\beta^{\prime-4}
\exp(-\nu t)\Big((1+\lambda\beta^{-1}(1-\exp(-\frac{\nu}{2}
t))\Big)^{-1},
\end{equation}
\begin{equation}
\partial_{t}{\cal W}=12\pi\lambda c^{-1}\kappa^{2}\beta^{\prime-4}
\exp(-\frac{\nu}{2} t)\end{equation} and
\begin{equation}
\partial_{t}S=12\pi\lambda\kappa^{2}c^{-1}\beta^{\prime-3}\exp(-\frac{\nu}{2}t)
\Big(1+\lambda\beta^{-1}\exp(-\frac{\nu}{2}t)\Big)^{-1},
\end{equation}
Eq.(67) means that the change of the entropy has the same sign as
the change of energy (66),i.e.,if the temperature of the system is
higher than the temperature of the reservoir then the entropy of
the system is decreasing, because the energy is flowing into the
reservoir. If the temperature of the reservoir is higher than the
temperature of the system then the entropy of the system is
increasing. The total change of the entropy of the system plus the
reservoir (the entropy production) is described by
\cite{mazur}\cite{glans}
 $-\partial_{t}{\cal F}=T\partial_{t}{\cal S}-\partial_{t}{\cal W}\geq
 0$.

 Note that the energy per mode
$w(t)$ (45) tends from its initial value

$w(t=0)=3\beta^{\prime-1}$ to the equilibrium limit $
w(t=\infty)=3\beta^{-1}$ in accordance with the equipartition
theorem for massless particles.

\section{Discussion}
We have extended the  diffusion  model of non-equilibrium
thermodynamics \cite{mazur} to relativistic diffusions. The model
describes a balance of entropy and energy: the exchange of the
entropy with the surroundings and the entropy production inside
the system. The  time derivative of the relative entropy (with an
opposite sign) is  interpreted as the entropy production in the
 total system consisting of the relativistic diffusing particles and the reservoir
 \cite{mazur}\cite{glans}\cite{nicolis}.  Such models can be useful
in a description of a beam (fluid) of relativistic particles
moving in a medium. As an example we could consider  a motion of
light through the space filled with electrons
 \cite{rybicki}\cite{komp} (this could include the initial stage
of the Big Bang \cite{zeld}); for a discussion of some other
applications of non-equilibrium radiation theory see \cite{rubin}.
We could consider also a diffusive motion of a heavy quark in the
plasma of gluons and light quarks. Such a model of diffusion (after
an extension to open systems) can be tested in the LHC experiments
\cite{ion}. The entropy exchange can play an important role in a
description of heavy ion collisions \cite{geiger}.

In this letter we have discussed a diffusion equation with a
probability distribution tending  to the classical relativistic
(J\"uttner) equilibrium distribution. We could consider in the
diffusion equations (2) the drifts leading to the quantum
equilibrium distributions. However, in such a case the
thermodynamics will change substantially, e.g., the entropy for
Bose-Einstein particles should have the form
\begin{displaymath}
S=-\int d{\bf x}d{\bf p}p_{0}\Phi\ln p_{0}\Phi+\int d{\bf x}d{\bf
p}(1+p_{0}\Phi)\ln(1+p_{0}\Phi)
\end{displaymath}
instead of eq.(39).  The classical statistical thermodynamics is
valid in the limit of large $\beta cp_{0}$. If this condition is
not satisfied then quantum phenomena appear. In the photon case
there are bunching (enhancement) effects resulting from
Bose-Einstein statistics described by non-linear terms in the
Boltzmann equation \cite{rybicki}\cite{komp} ( there is the Pauli
blocking factor in the case of the Fermi-Dirac statistics). Only
in the limit of low photon density these non-linear terms can be
neglected. A discussion of thermodynamics with quantum equilibrium
distributions requires non-linear relativistic diffusion equations
which will be studied in forthcoming publications.
\section{Appendix}
Let us assume that the initial state $\Phi_{E}^{\prime}$ is an
equilibrium state but with a temperature $T^{\prime}$ different
from the temperature $T$ of the reservoir
\begin{displaymath}
\Phi_{E}^{\prime}=\exp(-\beta^{\prime}cp_{0})
\end{displaymath}
 We are unable to obtain an exact time evolution of energy and
entropy of a stream of massive particles. However, we can obtain
their evolution for a small time calculating the time derivatives
at $t=0$ from eqs.(26) and (35). First, for the initial
distribution $\Phi_{E}^{\prime}$
\begin{displaymath}\begin{array}{l}
\partial_{t}{\cal W}(t=0)=\frac{3c^{2}\kappa^{2}}{2}Z-\frac{\kappa^{2}}{2} \beta c^{2} {\cal
W}_{0}+\frac{\kappa^{2}}{2}\beta m^{2}c^{5}\int d{\bf x}\rho
\cr=6\pi\kappa^{2}c^{-1
}\beta^{\prime-3}(1-\frac{\beta}{\beta^{\prime}})
K_{2}(mc^{2}\beta^{\prime})(mc^{2}\beta^{\prime})^{2}
\end{array}\end{displaymath} Then, after a calculation of
$\partial_{0}S_{K}(\Phi_{E}^{\prime},\Phi_{E})$ from eq.(35) we
obtain
\begin{displaymath}
\partial_{0}{\cal F}(t=0)=-6\pi\kappa^{2}c^{3}\beta^{-1}m^{2}
(1-\frac{\beta}{\beta^{\prime}})^{2} K_{2}(mc^{2}\beta^{\prime})
\end{displaymath}(where $K_{\nu}$ denotes the modified Bessel
function of order $\nu$). The limit $m\rightarrow 0$ of these
formulas agrees with eqs.(65)-(67) at $t=0$ as  $K_{2}(x)\simeq
2x^{-2}$ for a small $x$.


\begin{thebibliography}{99}\bibitem{pitaj} E.M. Lifshits and L.P. Pitaevskij,
Physical Kinetics,Pergamon Press,1981
\bibitem{mazur}S.R. de Groot and P.Mazur, Non-equilibrium Thermodynamics, North Holland,1969
 \bibitem{glans}P. Glansdorff and I. Prigogine, Thermodynamic
 Theory of Structure, Stability and
 Fluctuations, Wiley-Interscience,1971
 \bibitem{dunkel}J. Dunkel and P. H\"anggi,Phys.Rep.{\bf 471},1(2009)
\bibitem{debrev}C. Chevalier and F. Debbasch, AIP Conf.Proc.{\bf
913},42(2007)
\bibitem{rybicki}G.B. Rybicki and A.P. Lightman,

Radiative Processes in Astrophysics,Wiley-VCH,1979

\bibitem{zachar} P. Carruthers and F. Zachariasen,
Phys.Rev.{\bf D13},950(1976)
\bibitem{groot}S.R. de Groot, W.A. van Leeuwen and Ch.G. van Weert,  Relativistic
Kinetic Theory, North Holland,1980


\bibitem{hwa}R.C. Hwa,Phys.Rev.{\bf D32},637(1985)
\bibitem{svet} B.Svetitsky, Phys.Rev.{\bf D37},2484(1988)
\bibitem{rafelski}D.B. Walton and J. Rafelski,Phys.Rev.Lett.{\bf 84},31(2000)
\bibitem{ion}R.Rapp and H. van Hees,Int.Journ.Mod.Phys.{\bf E}(2009) arXiv:0903.1096



\bibitem{schay} G.Schay,PhD thesis,Princeton University,1961
\bibitem{dudley} R.Dudley, Arkiv for Matematik,{\bf 6},241(1965)
\bibitem{habapre}Z. Haba, Phys.Rev.{\bf E79},021128(2009)



\bibitem{habajmp}Z.Haba, ArXiv:0911.3126

\bibitem{landau} L.D. Landau and E.M. Lifshits,
Fluid Mechanics,Addison-Wesley,1958
\bibitem{israel}W.Israel, Ann.Phys.(N.Y.){\bf 100},310(1976)

\bibitem{rom}P. Romatschke, ArXiv:0902.3663

\bibitem{debentropy}M. Rigotti and F.Debbasch, Journ.Math.Phys.{\bf 46},103303(2005)

\bibitem{deb2}C. Chevalier and F. Debbasch, Mod.Phys.Lett.{\bf B22},383(2008)
\bibitem{haken}H. Haken and R. Graham, Zeitsch.Phys.{\bf
243},289(1971),{\bf 245},141(1971)
\bibitem{nicolis}D. Daems and G. Nicolis, Phys.Rev.{\bf
E59},4000(1999)
\bibitem{bag}B.Ch. Bag,S.K. Banik and D.S. Ray,Phys.Rev.{\bf
E64},026110(2001)
\bibitem{habampa}Z.Haba,Mod.Phys.Lett.{\bf A24},3193(2009)
\bibitem{komp}A.S. Kompaneets, Sov.Phys.-JETP,{\bf 4},730(1957)





\bibitem{kol}A.N. Kolmogorov,Math.Ann.{\bf 112},155(1936),{\bf
113},766(1937)

\bibitem{juttner}F. J\"uttner, Ann.Phys.(Leipzig){\bf 34},856(1911)
\bibitem{ikeda} N. Ikeda and S.
Watanabe, Stochastic Differential Equations and Diffusion
Processes,North Holland,1981

\bibitem{risken}H. Risken, The Fokker-Planck Equation,
Springer,1989








\bibitem{leb}M.S. Green, Phys.Rev.{\bf 20},1281(1952)
\bibitem{lebo} J.L. Lebowitz and P.G. Bergmann, Ann. Phys.(N.Y.){\bf
1},1(1957)
\bibitem{habajpa}Z. Haba,Journ.Phys.{\bf A42},445401(2009)
\bibitem{zeld}Ya.B. Zeldovich,Sov.Phys.-Usp.{\bf 18},79(1975)
\bibitem{rubin}A. Perez-Madrid,J.M.Rubi
and L.C. Lapas, ArXiv:1002.0794
\bibitem{geiger}K. Geiger, Phys.Rev.{\bf D46},4986(1992)





\end{thebibliography}
\end{document}